\documentclass[conference]{IEEEtran}

\usepackage[OT1]{fontenc}
\usepackage[numbers,sort&compress]{natbib}
\usepackage[cmex10]{amsmath}
\usepackage{amssymb}
\usepackage{bm}
\usepackage[dvipdfmx]{graphicx}
\usepackage{color}
\usepackage{subfigure}
\usepackage{tabularx}
\usepackage{arydshln}
\usepackage{mathtools}
\usepackage{multirow}
\usepackage{algorithm,algorithmic}
\usepackage{url}
\usepackage{listings}
\usepackage{comment}
\usepackage{physics}
\usepackage{braket}

\usepackage{hyperref}
\usepackage[absolute,overlay]{textpos}

\def\x{\mathbf{x}}
\def\b{\mathbf{b}}
\def\s{\mathbf{s}}
\def\y{\mathbf{y}}
\def\H{\mathbf{H}}
\def\A{\mathbf{A}}
\def\G{\mathbf{G}}
\def\ie{\textit{i.e.}}

\newcommand{\argmin}{\mathop{\rm arg~min}\limits}

\makeatletter
\def\ps@IEEEtitlepagestyle{%
  \def\@oddfoot{\mycopyrightnotice}%
  \def\@oddhead{\hbox{}\@IEEEheaderstyle\leftmark\hfil\thepage}\relax
  \def\@evenhead{\@IEEEheaderstyle\thepage\hfil\leftmark\hbox{}}\relax
  \def\@evenfoot{}%
}
\def\mycopyrightnotice{%
  \begin{minipage}{\textwidth}
  \centering \scriptsize
\textcopyright 2024 IEEE. Personal use of this material is permitted.
  Permission from IEEE must be obtained for all other uses, in any current or future
  media, including reprinting/republishing this material for advertising or promotional
  purposes, creating new collective works, for resale or redistribution to servers or
  lists, or reuse of any copyrighted component of this work in other works.
  \end{minipage}
}
\makeatother
\begin{document}
\title{Grover Adaptive Search for Maximum Likelihood Detection of Generalized Spatial Modulation\\[-1ex]}

\author{\IEEEauthorblockN{Kein Yukiyoshi\IEEEauthorrefmark{1}, Taku Mikuriya\IEEEauthorrefmark{1}, Hyeon Seok Rou\IEEEauthorrefmark{2}, Giuseppe Thadeu Freitas de Abreu\IEEEauthorrefmark{2}, and Naoki Ishikawa\IEEEauthorrefmark{1}}
\IEEEauthorblockA{\IEEEauthorrefmark{1}Graduate School of Engineering Science, Yokohama National University, 240-8501 Kanagawa, Japan.\\
\IEEEauthorrefmark{2}School of Computer Science and Engineering, Constructor University, Campus Ring 1, 28759 Bremen, Germany.\\[-3ex]
}}

%

\maketitle
\TPshowboxesfalse
\begin{textblock*}{\textwidth}(45pt,10pt)
\footnotesize
\centering
Accepted for presentation at the IEEE 100th Vehicular Technology Conference (VTC2024-Fall).\\
This is the author's version which has not been fully edited and content may change prior to final publication.
\end{textblock*}

\begin{abstract}
We propose a quantum-assisted solution for the maximum likelihood detection (MLD) of generalized spatial modulation (GSM) signals. 
Specifically, the MLD of GSM is first formulated as a novel polynomial optimization problem, followed by the application of a quantum algorithm, namely, the Grover adaptive search.
The performance in terms of query complexity of the proposed method is evaluated and compared to the classical alternative via a numerical analysis, which reveals that under fault-tolerant quantum computation, the proposed method outperforms the classical solution if the number of data symbols and the constellation size are relatively large.
\end{abstract}


\IEEEpeerreviewmaketitle

\vspace{-1ex}
\section{Introduction}

Triggered by the study conducted in \cite{mesleh2006spatial}, spatial modulation (SM) has been extensively researched as a potential solution to address the fundamental trade-off between performance and complexity in multiple-input multiple-output (MIMO) systems.
In SM systems, information bits are conveyed not only via data symbols but also through the activation pattern (AP) of transmit antennas in each time slot.
This dual-layer modulation enables high spectral efficiency while maintaining low computational complexity at the receiver \cite{ishikawa201850}.

In the original SM scheme, only a single transmit antenna is activated within a symbol duration, whereas generalized SM (GSM) extends the approach by simultaneously activating an integer number $K$ of antennas, thus achieving higher transmission rates \cite{jeganathan2008generalized}, albeit at the expense of higher detection complexity.
And although various lower complexity methods for GSM schemes exist \cite{tang2013new, An_TVT22,Rou_Asilomar22_QSM,rou2022scalable,Rou_CAMSAP23}, the reducing in detection complexity unavoidably implicates in the degradation of other performance metrics.

As Moore's law nears its end and the limits of transistor miniaturization are approached, the performance of classical computation is anticipated to plateau.
In fact, even if Moore's law continues to hold for another decade \cite{irds2022international}, it is expected that quantum computers will overcome current limitations and supersede classical computers, at least for a selected number of tasks \cite{choi2023quantum}, before such a time span.
To offer some examples, the well-known quantum annealing (QA) \cite{kadowaki1998quantum} and quantum approximate optimization algorithm (QAOA) \cite{farhi2014quantum} are two heuristic approaches that have already been demonstrated to work effectively on real devices.
And although both approaches have proven unable to outperform classical computation under realistic assumptions \cite{stilckfranca2021limitations}, mechanisms such as fault-tolerant quantum computation (FTQC) \cite{suzuki2022quantum} are also expected to continue developing which will resolve such shortcomings.

Grover adaptive search (GAS) is known as a promising quantum algorithm designed for FTQC devices, which is capable of solving a certain class of optimization problems with a quadratic speedup in query complexity \cite{durr1999quantum}.
Recently, a method for constructing quantum circuits of GAS for polynomial binary optimization problems was proposed and tested on a real quantum computer \cite{gilliam2021grover}.
Owing to its versatile capabilities and algorithmic efficiency, GAS has been applied to various types of problems in wireless communications, including general MIMO MLD \cite{norimoto2023quantum}, joint MLD of power-domain non-orthogonal multiple access (NOMA) systems \cite{norimoto2023grover}, channel assignment problems \cite{sano2023qubit}, and the construction of binary constant weight codes \cite{yukiyoshi2022quantum}.

Despite the significant improvements in efficiency demonstrated using GAS across multiple domains, to the best of our knowledge, its application in SM detection remains unexplored, which is noticeable as GAS is a vital approach to reduce complexity while maintaining the MLD performance.
Against this background, we offer in this paper a new formulation a MLD of GSM as a polynomial optimization problem.
Using this novel formulation, we efficiently solve the problem by applying GAS, and evaluate its query complexity in comparison with the classical exhaustive search method through numerical analysis.

\section{Grover Adaptive Search for Polynomial Binary Optimization}
\label{sec:GAS}

In this section, we review and contextualize GAS \cite{durr1999quantum}, a quantum algorithm designed to obtain the optimal solution of an optimization problem, whose query complexity is $\mathcal{O}\big(\sqrt{N / t}\,\big)$\footnote{$\mathcal{O}(\cdot)$ denotes the big-O notation~\cite{knuth1976big}.}, where $N$ is the search space size and $t$ is the number of solutions.
%
%
While the original GAS algorithm assumes the existence of a black-box quantum oracle $\mathbf{O}$ in order to identify the desired states \cite{durr1999quantum}, such an assumption was partially alleviated in \cite{gilliam2021grover}, where an efficient method was proposed to construct concrete quantum circuits for the polynomial binary optimization problems in the form
\begin{align}
\label{eq:PolynomialOptProb}
\min_{\x} \quad & E(\x) \\
\textrm{s.t.} \quad & x_i \in \mathbb{F}_2 = \{0, 1\} ~~~~ (i = 0, \cdots, n-1),
\nonumber
\end{align}
where $\x = (x_0, \cdots\!, x_{n - 1})$ denotes binary variables and $E(\x) \in \mathbb{Z}$ denotes an arbitrary polynomial objective function.

Gilliam's construction method \cite{gilliam2021grover} requires $n + m$ qubits, where $n$ and $m$ respectively denote the number of binary variables and the number of qubits required to encode $E(\x)$.

\begin{algorithm}[H]
\caption{Procedure of GAS~\cite{gilliam2021grover,norimoto2023quantum}\label{alg:GAS}}
\begin{algorithmic}[1]
\renewcommand{\algorithmicrequire}{\textbf{Input:}}
\renewcommand{\algorithmicensure}{\textbf{Output:}}
\REQUIRE $E:\mathbb{F}_2^{n}\rightarrow\mathbb{Z}, \lambda=8/7$
\ENSURE $\x$
\STATE {Uniformly sample $\x_0 \in \mathbb{F}_2^n$ and set $y_0=E(\x_0)$}.
\STATE {Set $k = 1$ and $i = 0$}.
\REPEAT
\STATE\hspace{\algorithmicindent}{Randomly select the rotation count $L_i$ from the set $\{0, 1, ..., \lceil k-1 \rceil$\}}.
\STATE\hspace{\algorithmicindent}{Evaluate $\G^{L_i} \A_{y_i} \Ket{0}_{n + m}$ to obtain $\x$ and $y = E(\x)$}.
\vspace{-3ex}
\hspace{\algorithmicindent}\IF{$y<y_i$}
\STATE\hspace{\algorithmicindent}{$\x_{i+1}=\x, y_{i+1}=y,$ and $k=1$}.
\hspace{\algorithmicindent}\ELSE{\STATE\hspace{\algorithmicindent}{$\x_{i+1}=\x_i, y_{i+1}=y_i,$ and $k=\min{\{\lambda k,\sqrt{2^n}}\}$}}.
\ENDIF
\STATE{$i=i+1$}.
\UNTIL{a termination condition is met}.
\end{algorithmic}
\end{algorithm}

Expressing $ E(\x)$ via the two's complement representation, it can be seen that $m$ must satisfy
\vspace{-0.5ex}
\begin{equation}
-2^{m - 1} \leq \min_{\x} E(\x) \leq \max_{\x} E(\x) < 2^{m - 1}.
\end{equation}

In view of the above, the GAS procedure, which is summarized in the form of a pseudo-code in Algorithm \ref{alg:GAS}, can be described as follows\footnote{Open-source implementation available from IBM Qiskit~\cite{qiskit}.}.
First, an initial threshold $y_0 = E(\x_0)$ is set by using a random input $\x_0$.
Then, a rotation count $L_i$ is randomly drawn from the uniform distribution $[0, k)$, where the parameter $k$ is increased at each iteration by the rule $k_{i+1} = \min\qty{\lambda k, \sqrt{2^n}}$.
Using the drawn rotation count $L_i$, we generate and measure a specified quantum state to obtain a better solution $\x$ such that the value of the objective function is less than the current threshold $y_{i}$, and subsequently update the threshold via $y_{i+1} = E(\x)$.
The optimal solution is obtained through multiple measurements of the quantum states.

\begin{figure}[H]
\includegraphics[width=\columnwidth]{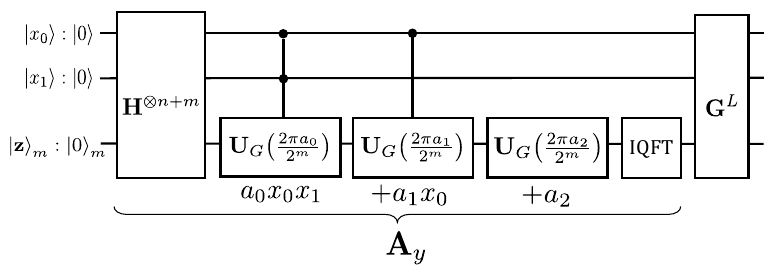}
\vspace{-4ex}
\caption{Quantum circuit of the GAS corresponding to problem \eqref{eq:PolynomialOptProb} with the objective function $E(\mathbf{x}) - y = a_0x_0x_1 + a_1x_0 + a_2$.}
\label{fig:circuit1}
\vspace{-1ex}
\end{figure}

Next, we describe the corresponding circuit construction method preparing the quantum state $\G^{L_i} \A_{y} \Ket{0}_{n+m}$~\cite{gilliam2021grover}.
Here, $\A_{y}$ represents the state preparation operator that encodes the values of $E(\x) - y$ into a quantum superposition for arbitrary inputs $\x$, and $\G$ denotes the Grover operator, which is used to amplify the amplitude of states $\x$ satisfying $E(\x) - y < 0 \Leftrightarrow E(\x) < y$.
As an example, Fig.~\ref{fig:circuit1} shows a quantum circuit constructed to represent the objective function $E(\mathbf{x}) - y = a_0x_0x_1 + a_1x_0 + a_2$, where $a_2$ is a constant that includes both the constant term of $E(\mathbf{x})$ and $-y$.
Each polynomial term $a_0x_0x_1$, $a_1x_0$, and $a_2$ is represented by a corresponding quantum gate $\mathbf{U}_{G}(\theta)$.
The construction involves the following steps.

\noindent {\bf 1) Hadamard gates:} Applying Hadamard the gates $\H^{\otimes (n+m)}$ to the initial state $\Ket{0}_{n+m}$ transforms it into a uniform superposition state.
This transition is described by \cite{gilliam2021grover}
\begin{eqnarray}
\label{eq:AyH}
\Ket{0}_{n+m}\; \xrightarrow{\H^{\otimes (n+m)}} \hspace{-3ex}&& \frac{1}{\sqrt{2^{n+m}}} \sum_{i=0}^{2^{n+m}-1} \Ket{i}_{n+m} =\\
&&\frac{1}{\sqrt{2^{n+m}}} \sum_{\x \in \mathbb{F}_2^{n}} \sum_{i=0}^{2^{m}-1} \Ket{\x}_{n} \Ket{i}_{m},\nonumber
\end{eqnarray}

\noindent where the Hadamard gates $\H^{\otimes (n+m)}$ is obtained using the tensor product $\otimes$, defined as
\begin{equation}
\H^{\otimes (n+m)} = \underbrace{\H \otimes \H \otimes \cdots \otimes \H}_{n+m},
\end{equation}
with
\begin{equation}
\H = \frac{1}{\sqrt{2}}\mqty[1 & 1 \\ 1 & -1].
\end{equation}

\noindent {\bf 2) Thresholding:} Each term of the difference $E(\x) - y$ comparing the objective function $E(\x)$ with the threshold $y$ corresponds to a specific unitary gate $\mathbf{U}_{G}(\theta)$, which acts on the lower $m$ qubits and rotates the phase of quantum states. Given a coefficient of the term $a$, the phase angle is given by $\theta = \frac{2 \pi a}{2^{m}}$. The unitary gate is then defined by
\begin{equation}
\mathbf{U}_{G}(\theta) = \underbrace{\mathbf{R}(2^{m - 1}\theta) \otimes \cdots \otimes \mathbf{R}(2^{0}\theta)}_{m},
\end{equation}
where the phase gate $\mathbf{R}(\theta)$ is defined by
\begin{equation}
\mathbf{R}(\theta) = \mqty[1 & 0 \\ 0 & e^{j\theta}].
\end{equation}

For a state $\x'$, where $\theta' = 2 \pi \qty(E(\x') - y) / 2^{m}$, the application of $\mathbf{U}_{G}(\theta')$ yields the transition of~\cite{gilliam2021grover}
\begin{equation}
\frac{1}{\sqrt{2^{m}}} \sum_{i=0}^{2^{m}-1} \Ket{i}_m \xrightarrow{\mathbf{U}_{G}\qty(\theta')}
\frac{1}{\sqrt{2^m}}\sum_{i=0}^{2^{m} - 1}e^{ji\theta'} \Ket{i}_m.
\label{eq:AyUG}
\end{equation}

\noindent {\bf 3) Inverse Quantum Fourier Transform (IQFT):} The IQFT \cite{shor1997polynomialtime} acts on the lower $m$ qubits, yielding the transition of \cite{gilliam2021grover}
\begin{equation}
\frac{1}{\sqrt{2^m}}\sum_{i=0}^{2^m - 1}e^{ji\theta'} \Ket{i}_m  \xrightarrow{\mathrm{IQFT}} \frac{1}{\sqrt{2^m}}\Ket{E(\mathbf{x}_0) - y}_m.
\end{equation}

These three steps described above are collectively termed as the \emph{state preparation operator} $\A_{y}$, which encodes the value $E(\mathbf{x}) - y$ into $m$ qubits for arbitrary $\x \in \mathbb{F}_2^{n}$, satisfying \cite{gilliam2021grover}
\begin{equation}
\A_{y} \Ket{0}_{n+m} = \frac{1}{\sqrt{2^n}}\sum_{\x \in \mathbb{F}_2^n} \Ket{\x}_{n}\Ket{E(\mathbf{x})-y}_{m}.
\end{equation}

\noindent {\bf 4) Identification:} Finally, to amplify the desired states, the Grover operator $\G = \A_{y}\mathbf{D}\A_{y}^H\mathbf{O}$ is applied $L$ times, where $\mathbf{D}$ denotes the Grover diffusion operator \cite{grover1996fast}.
The oracle $\mathbf{O}$ identifies the desired states satisfying $E(\x) - y < 0$. Given that we utilize the two's complement representation, such states can be identified by a single Pauli-Z gate shown below, acting on the most significant of the $m$ qubits
\begin{equation}
\mathbf{Z} = \mqty[1 & \;\;0 \\ 0 & -1].
\end{equation}

While the coefficients of the objective function are typically restricted to integers, this circuit construction method can also accommodate real-valued coefficients at the cost of the accurate encoding of the objective function~\cite{gilliam2021grover,norimoto2023quantum}. This instability can be ignored by assuming a sufficiently large number $m$ of qubits.

\section{System Model and Classical MLD}

Consider an $N_t\times N_r$ multiple-input multiple-output (MIMO) system and assume independent and identically distributed (i.i.d.) frequency-flat Rayleigh fading channels, such that each element of the channel matrix $\H \in \mathbb{C}^{N_r \times N_t}$ and the noise vector $\mathbf{n} \in \mathbb{C}^{N_r \times 1}$ follows the complex Gaussian distributions $\mathcal{CN}(0, 1)$ and $\mathcal{CN}(0, \sigma_n^2)$, respectively.
Given a noise variance $\sigma_n^2$ and a transmission power $\sigma_s^2$, the corresponding SNR is then defined as $\rho = \sigma_s^2/\sigma_{n}^2$.

Let $s_k$ be the $k$-th element of the data symbol vector $\s \in \mathbb{C}^{K \times 1}$, and let $L$ be the constellation size. Each symbol $s_k$ is derived from a corresponding bit sequence, $\b_\s^{(k)} = (b_\s^{(k, 0)}, \cdots, b_\s^{(k, L-1)})$, using a mapping function $s(\b_\s^{(k)})$ specified in the 5G NR standard~\cite{3gpp2018ts}.
For example, in the case of binary phase-shift keying (BPSK), the mapping function is specified as
\begin{equation}
\label{eq:bpsk}
s_{\mathrm{BPSK}}(\b_\s^{(k)}) = \frac{1}{\sqrt{2}}[(1 - 2b_\s^{(k, 0)}) + j(1 - 2b_\s^{(k, 0)})],
\end{equation}
where $j = \sqrt{-1}$ denotes the imaginary unit, while
for the case of quadrature PSK (QPSK), we have
\begin{equation}
\label{eq:qpsk}
s_{\mathrm{QPSK}}(\b_\s^{(k)}) = \frac{1}{\sqrt{2}}[(1 - 2b_\s^{(k, 0)}) + j(1 - 2b_\s^{(k, 1)})],
\end{equation}
and finally, for the case of 16-QAM, we have
\begin{equation}
\label{eq:16qam}
\begin{split}
s_{\mathrm{16QAM}}(\b_\s^{(k)}) = &\frac{1}{\sqrt{10}} (1 - 2b_\s^{(k, 0)})[2 - (1 - 2b_\s^{(k, 2)}]\\
+ &j\frac{1}{\sqrt{10}} (1 - 2b_\s^{(k, 1)})[2 - (1 - 2b_\s^{(k, 3)}].
\end{split}
\end{equation}

Next, let $a_{(i, k)}$ be the $(i, k)$ element of an antenna activation patter (AP) matrix $\A$, such that a symbol $s_k$ is transmitted via $i$-th antenna when $a_{(i, k)} = 1$, with the corresponding GQSM codeword $\mathbf{c} \in \mathbb{C}^{N_t \times 1}$ given by $\mathbf{c} = \A\s$.

\begin{table}[tb]
\centering
\caption{AP Mapping Table For $N_t = 4$, $K = 3$ and $Q = 4$}
\vspace{-2ex}
\begin{tabular}{ccc}
\hline
Bit sequence $\b_\A$ & Antenna indices & AP matrix $\A$\\
\hline
$00$ & $(0, 1, 2)$ & $\mqty[1 & 0 & 0 \\ 0 & 1 & 0 \\ 0 & 0 & 1 \\ 0 & 0 & 0]$\\
$01$ & $(1, 2, 3)$ &$\mqty[0 & 0 & 0 \\ 1 & 0 & 0 \\ 0 & 1 & 0 \\ 0 & 0 & 1]$ \\
$10$ & $(2, 3, 0)$ & $\mqty[0 & 0 & 1 \\ 0 & 0 & 0 \\ 1 & 0 & 0 \\ 0 & 1 & 0]$ \\
$11$ & $(3, 0, 1)$ & $\mqty[0 & 1 & 0 \\ 0 & 0 & 1 \\ 0 & 0 & 0 \\ 1 & 0 & 0]$ \\
\hline
\end{tabular}
\label{tab:mapping}
\end{table}

The codebook of APs is given by
\begin{equation}
\mathbb{A} = \bigl\{ \A \in \qty{0, 1}^{N_t \times K} \mid \forall k = 1, \cdots K,\ \Sigma_{i=1}^{N_t} a_{(i, k)} = 1 \bigr\},
\end{equation}
where we emphasize that the number of APs, $i.e.$ $Q = |\mathbb{A}|$, needs to be a power of 2 and satisfies a constraint $2 \leq Q \leq 2^{\lfloor\log_2\binom{N_t}{K}\rfloor} \leq \binom{N_t}{K}$.

Notice that $Q$ is not limited to its maximum possible value; it can be a lower number to achieve additional gain at the expense of a reduced transmission rate \cite{ishikawa2019imtoolkit}.
An AP matrix $\A$ is derived from a corresponding bit sequence, $\b_\A = (b_\A^{(0)}, \cdots, b_\A^{(\log_2 Q)})$, using a predefined mapping table, as exemplified in Table \ref{tab:mapping}.

From the above, it follows that the received signal $\y$ can be modeled as
\begin{equation}
\label{eq:system}
\y = \H\A\s + \mathbf{n} \in \mathbb{C}^{N_r \times 1},
\end{equation}
such that the corresponding MLD can be formulated as
\begin{equation}
\label{eq:MLD}
(\hat{\s}, \hat{\A}) = \underset{\s, \A}{\argmin} \norm{\y - \H\A\s}_{\mathrm{F}}^2,
\end{equation}
where $\norm{\cdot}_\mathrm{F}$ denotes the frobenius norm.

The estimated bit sequence $(\hat{\b}_\s, \hat{\b}_\A)$ is obtained through an inverse mapping operation for $(\hat{\s}, \hat{\A})$.
Note that, while several low-complexity detectors have been developed for SM schemes  \cite{tang2013new, An_TVT22,Rou_Asilomar22_QSM,rou2022scalable,Rou_CAMSAP23}, optimality is generally guaranteed only by performing an exhaustive search over MLD.

\section{Proposed Quantum-Assisted MLD of GSM}

In this section, we first formulate the MLD of GSM as a binary optimization problem, aiming to obtain a solution through GAS.
We then evaluate the computational complexity of our proposed GAS-based method, as well as the classical exhaustive search alternative, which are used to perform a numerical analysis to compare them.

\subsection{Proposed Problem Formulation}
\label{subsec:prob_form}

By dividing the binary variables $\x$ into those representing data symbols $\x_\s = (\x_\s^{(0)}, \cdots, \x_\s^{(N_t - 1)})$, and those associated with AP $\x_\A = (x_\A^{(0)}, \cdots, x_\A^{(N_t - 1)})$, the MLD of GSM can be formulated as a polynomial binary optimization problem with the objective function
\begin{eqnarray}
\label{eq:obj_func}
E(\x_\s, \x_\A) = \norm{\y - \H\mathbf{c}(\x_\s, \x_\A)}_{\mathrm{F}}^2 && \\
&&\hspace{-25ex} + \lambda_1 \qty(\sum_i^{N_t} x_{\A}^{(i)} - K)^2 + \lambda_2 \sum_{\A \notin \mathbb{A}} \qty(\prod_{i \in \A} x_{\A}^{(i)}),
\nonumber
\end{eqnarray}
where $\mathbf{c}(\x_\s, \x_\A)$ represents a GSM codeword given by
\begin{equation}
\label{eq:codeword}
\mathbf{c}(\x_\s, \x_\A) = \mqty[s(\x_\s^{(0)}) x_\A^{(0)} &
\cdots & s(\x_\s^{(N_t - 1)}) x_\A^{(N_t - 1)}]^\mathrm{T}\!\!\!\!,\hspace{-1ex}
\end{equation}
with each symbol $s(\x_\s^{(i)})$ in \eqref{eq:codeword} constructed via a mapping function such as those in equations \eqref{eq:bpsk}, \eqref{eq:qpsk} and \eqref{eq:16qam}.

We clarify that the first term in equation \eqref{eq:obj_func} is for finding the codeword with the maximum likelihood corresponding to \eqref{eq:MLD}, while the second and third terms are penalty functions that constraints the number of activated antennas to $K$, and enforces an exclusion of AP not included in the predefined APs $\mathbb{A}$, respectively.

\subsection{Complexity Analysis}
\label{subsec:complexity}

As described in Section \ref{sec:GAS}, the GAS offers a quadratic speedup for binary optimization problems.
Since equation \eqref{eq:obj_func} contains $n = N_t\log_2(L) + N_t$ binary variables, it follows that the query complexity of the proposed method for MLD of GSM is given by
\begin{equation}
\label{eq:comp_GAS}
f = \mathcal{O}(\sqrt{2^n}) = \mathcal{O}\qty(\sqrt{2^{N_t\log_2(L) + N_t}}).
\end{equation}

In contrast, the computational complexity required by the classical MLD via exhaustive search as per equation \eqref{eq:MLD}, is proportional to two raised to the power of the transmission rate $R$, which is represented by
\begin{equation}
\label{eq:comp_classic}
g = \mathcal{O}(2^R) = \mathcal{O}(L^KQ).
\end{equation}

\begin{figure}[tb]
\centering
\includegraphics[keepaspectratio,scale=0.69]{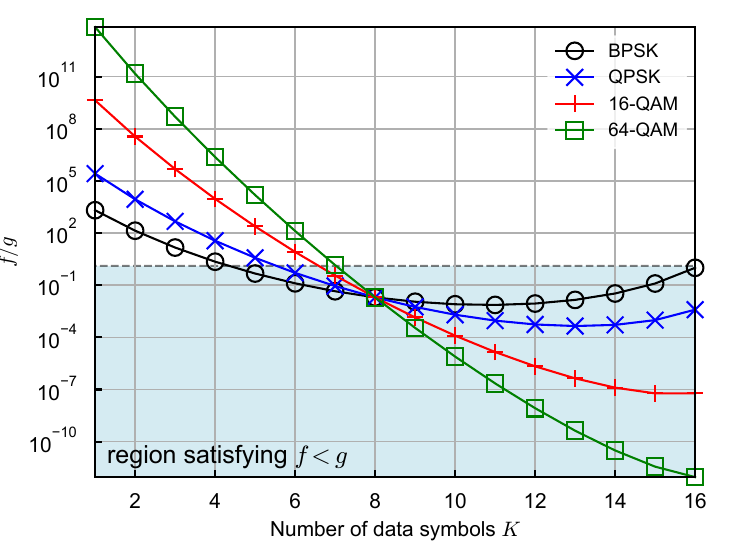}
\caption{Complexity ratio $f/g$, with $N_t = 16$ and $Q = \binom{N_t}{K}$.}
\label{fig:complexity_comparison}
\end{figure}

The complexities $f$ and $g$ are quite different, and their ratio is highly dependent on the actual values of the parameters involved.
To illustrate, Fig.~\ref{fig:complexity_comparison} shows the complexity ratio $f/g = \sqrt{2^{N_t\log_2(L) + N_t}} / (L^KQ)$ under different conditions.
It can be seen that the proposed method is superior ($i.e.$, has lower complexity) as the number of data symbols $K$ and the constellation size $L$ increase.
Ideally, the application of GAS reduces the complexity to the square root of the original search space.
However, since the proposed formulation introduces additional variables, a pure quadratic speedup cannot be achieved.
Specifically, equation \eqref{eq:codeword} includes $N_t$ symbol candidates, while the actual number of the transmitted data symbols is $K$.
This redundancy arises from the sparse structure of SM.
To the best of our knowledge, however, no alternative method existing to formulate this type of problem as a binary optimization without expanding the search space.

It is important to note that this comparison does not promise improvements in the actual time required for detection, since the concepts of time complexity in classical computing and query complexity in quantum computing are distinct.
Furthermore, this analysis does not take into account the constant factors included in the Big-O notation.
However, in the absence of a scalable FTQC, we consider this comparative approach to be at least informative as a reference.

\section{Performance Results}

In this section, we investigate the query complexity of the proposed quantum-assisted method, with the results of the classical exhaustive search method are also presented as references.
It is assumed that a sufficiently large number of qubits $m$ is available, and the following system parameters\footnote{Since $Q = \binom{N_t}{K}$, the APs $\mathbb{A}$ includes all the possible AP. Therefore the third term in \eqref{eq:obj_func} is not necessary and $\lambda_2$ is undefined here.} are utilized: $N_t = 4$, $N_r = 4$, $K = 3$, $Q = 4$, $\lambda_1 = 15$, and SNR $\rho = 0\ \mathrm{dB}$ with QPSK.
All results are averaged over $10^3$ independent channel realizations and two performance metrics adopted in \cite{botsinis2014fixedcomplexity}, namely, the query complexity in the quantum domain (QCQD), defined as the total number of Grover operators, \ie, $\sum_i L_i$; and the query complexity in the classical domain (QCCD), defined as the number of measurements of the quantum states, are considered.
We emphasize that the QCQD is as important to quantum algorithms as time complexity is to classical algorithms, and that since GAS performs an exhaustive search over the search space, both GAS and classical exhaustive search achieve optimal performance.
\begin{figure}[tb]
\centering
\includegraphics[width=\columnwidth]{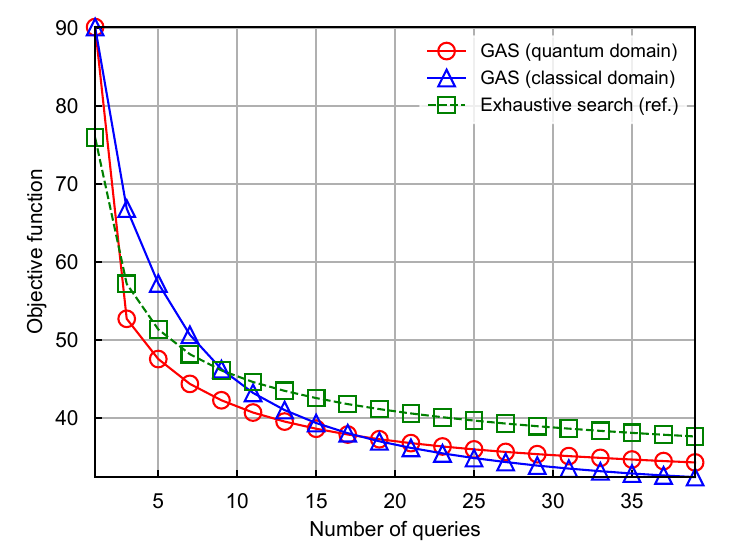}
\vspace{-2ex}
\caption{Relationship between the query complexity and the average of the objective function values.}
\label{fig:iter_objective}
\end{figure}
\begin{figure}[tb]
\includegraphics[width=\columnwidth]{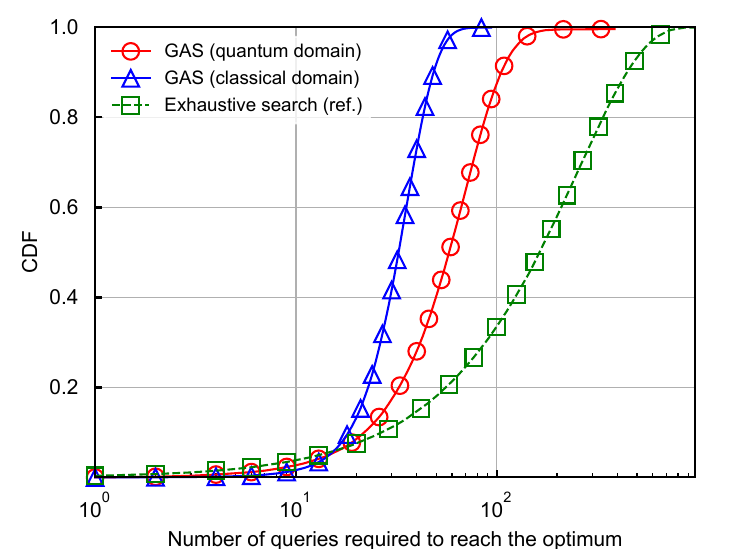}
\vspace{-2ex}
\caption{Cumulative distribution function of the query complexity required to reach the optimal solution.}
\label{fig:iter_cdf}
\vspace{-1ex}
\end{figure}

First, Fig. \ref{fig:iter_objective} shows the relationship between the query complexity and the average of the objective function values.
It is found that the proposed GAS-base method demonstrates superior convergence compared to the exhaustive search method in both quantum and classical domains.
Notice that in regions with a smaller number of queries, the proposed method results in higher objective function values than the exhaustive search method, which is caused by the proposed problem formulation incorporating additional redundant search space, as discussed in Section~\ref{subsec:complexity}, which leads to higher objective function values violating the penalty terms in \eqref{eq:obj_func}.

Given that GAS is a nondeterministic algorithm, its probabilistic performance is investigated in Fig. \ref{fig:iter_cdf}, which shows the cumulative distribution function of the query complexity required to reach the optimum.
As can be seen, the proposed method finds the optimal solution with lower query complexity in terms of both average and worst-case scenarios.

\section{Conclusions}
\label{sec:conc}

We introduced a quantum-assisted method for MLD of GSM, that finds optimal solutions while reducing the query complexity.
In order to apply GAS, the MLD of GSM was formulated as a new polynomial optimization problem.
Numerical analysis showed that the proposed method has a lower query complexity than the classical MLD with exhaustive search, as the number of data symbols and the constellation size increase; and that  the proposed method converges fast to the optimum in both quantum and classical domains.
The results motivates further investigation of for quantum speedup in future applications.

\footnotesize{
	\bibliographystyle{IEEEtranURLandMonthDiactivated}
	\bibliography{main}
}

\end{document}